\documentclass%
[amssymb,aps,prd,amsmath,floatfix,twocolumn,%
showpacs,nofootinbib,superscriptaddress]%
{revtex4}

\usepackage{graphicx}
\usepackage{color}

\begin{document}

\graphicspath{{Plots/}}

\title{ Testing the Accuracy and Stability of Spectral Methods in
  Numerical Relativity }

\author{Michael Boyle}
\author{Lee Lindblom}
\author{Harald P. Pfeiffer}
\author{Mark A. Scheel}
\affiliation{ 
  Theoretical Astrophysics 103-33,
  California Institute of Technology,
  Pasadena, CA, 91125
}

\author{Lawrence E. Kidder}
\affiliation{
  Center for Radiophysics and Space Research, 
  Cornell University, 
  Ithaca, NY, 14853
}

\date{\today}

\begin{abstract}
  The accuracy and stability of the Caltech-Cornell pseudospectral
  code is evaluated using the Kidder, Scheel, and Teukolsky (KST)
  representation of the Einstein evolution equations.  The basic
  ``Mexico City Tests'' widely adopted by the numerical relativity
  community are adapted here for codes based on spectral methods.
  Exponential convergence of the spectral code is established,
  apparently limited only by numerical roundoff error or by truncation
  error in the time integration.  A general expression for the growth
  of errors due to finite machine precision is derived, and it is
  shown that this limit is achieved here for the linear plane-wave
  test.
\end{abstract}


\maketitle


\section{Introduction}
\label{sec:Introduction}

A number of groups have now developed numerical relativity codes
sophisticated enough to evolve binary black-hole
spacetimes~\cite{Pretorius2005a,Baker2006,Campanelli2006,Diener2006,
  Herrmann2006,Scheel2006}.  The gravitational waveforms predicted by
these evolutions will play an important role in detecting and
interpreting the physical properties of the sources of these waves
soon to be detected (we presume) by the community of gravitational
wave observers (e.g., LIGO, etc.).  Therefore, such codes must be
capable of performing stable and accurate simulations of very
nonlinear and dynamical spacetimes.

Several years ago a large subset of the numerical relativity
community, the ``Apples with Apples''
collaboration~\cite{Alcubierre2004}, proposed a series of basic code
tests designed to verify the accuracy, stability, robustness and
efficiency of any code designed to find fully three-dimensional
solutions to the Einstein evolution equations.  These tests---often
referred to as the ``Mexico City Tests'' because they were first
formulated during a conference in Mexico City in May 2002---were
designed to be analogous to the standard suite of tests used by the
numerical hydrodynamics community (e.g., tests to reproduce Sedov
explosions, Sod shock tubes, blast waves, etc.) to commission new
hydrodynamics codes.  The Mexico City tests were designed to be
applicable to any formulation of Einstein's equations solved with any
numerical method.  All tests proposed so far concern bulk properties
of the formulation and numerical method, and so all of the evolutions
are carried out on a numerical grid with three-torus topology; no
boundary conditions are needed (or tested).  There are four basic
tests, some of them in a number of variations: (a) the evolution of
certain ``random'' initial data; (b) the evolution of small-amplitude
``linear'' plane-wave initial data; (c) the evolution of a nonlinear
gauge-wave representation of flat spacetime; and (d) the evolution of
initial data for a very dynamic and nonlinear Gowdy cosmological
model.

The Mexico City tests have now been applied to a number of different
numerical relativity codes that use different formulations of the
Einstein equations~\cite{Alcubierre2004,Jansen2006}.  But all of the
codes tested so far use finite difference numerical methods.  In this
paper we report the results of applying these tests to the code being
developed by the collaboration between the Caltech and Cornell
numerical relativity groups.  We use a first-order
symmetric-hyperbolic formulation of the equations developed by Kidder,
Scheel and Teukolsky~\cite{Kidder2001} (sometimes referred to as the
KST formulation) and we solve the equations using pseudospectral
numerical methods.  The results reported here differ therefore from
all previously tested cases both in the formulation of the Einstein
equations and the numerical methods used to solve them.

This paper is organized as follows: In
Sec.~\ref{sec:AdaptingTheMexicoCityTests} we review the KST
formulation of the equations and the pseudospectral numerical methods
that we use to solve them.  The remaining sections present the results
of the various Mexico City tests, adapted somewhat to provide more
challenging tests of a code based on spectral methods.  In
Sec.~\ref{sec:RandomInitialData} we show that our code is stable when
evolving small random perturbations of flat spacetime.  In
Sec.~\ref{sec:LinearWave} we report the results of the small-amplitude
plane-wave test.  We demonstrate the convergence rates for different
spatial resolutions and different time-step algorithms.  We also
derive an equation for the error introduced by finite machine
precision, and show that it limits the convergence of our evolutions
for small spatial and temporal resolutions.  In
Sec.~\ref{sec:GaugeWave} we show the stability of our evolution code
for nonlinear gauge waves.  In this case, nonlinear terms give rise to
an instability that is drastically reduced by suitably filtering the
components of the spectral expansion.
Section~\ref{sec:GowdySpacetime} shows the performance of our code for
evolutions of the highly dynamical Gowdy spacetime, in which the exact
analytical expressions for the components of the fields grow
exponentially in time.  Finally, we discuss and summarize our various
results in Sec.~\ref{sec:conclusion}.


\section{Solution Method}
\label{sec:AdaptingTheMexicoCityTests}

In this section we describe the formulation of the Einstein equations
and the pseudospectral numerical solution method that we test.  The
Mexico City tests were designed with finite-difference methods in mind
and were originally applied to formulations of the Einstein equations
that are second-order in space and first-order in time.  Both our
numerical methods and our representation of the Einstein equations
differ significantly from those in Ref.~\cite{Alcubierre2004}, so
appropriate modifications to the Mexico City test suite (for example,
the number of grid points used or the constraint quantities observed)
are needed.  These modifications are also described in this section.

\subsection{KST Formulation}
\label{subsec:KSTFormulation}

The KST system~\cite{Kidder2001} is a first-order symmetric hyperbolic
generalization of York's representation of the ADM
equations~\cite{york79}.  The dynamical variables of this system are
the three-metric $g_{ij}$, the extrinsic curvature $K_{ij}$, and a new
variable $D_{kij}$ that is initially set equal to $\partial_k g_{ij} /
2$. This last variable allows the system to be put into first-order
form.  Its introduction results in two additional constraints:
\begin{align}
  C_{kij} &\equiv D_{kij} - \frac{1}{2}\, \partial_k g_{ij}\ ,
  \label{eq:ThreeIndexConstraint}\\
  C_{lkij} &\equiv \partial_{[l} D_{k]ij}\ .
  \label{eq:FourIndexConstraint}
\end{align}
The KST evolution equations are obtained from the ADM
equations~\cite{york79} by adding constant multiples of the various
constraints to the evolution equations and by replacing the lapse with
a lapse density function.  These changes do not affect the physical
solutions of the system, but they do modify the unphysical
constraint-violating solutions.  The added constraint terms are
proportional to constant parameters $\left\{ \gamma_1, \gamma_2,
  \gamma_3, \gamma_4 \right\}$, which are chosen to make the system
symmetric hyperbolic~\cite{Kidder2005}.  The principal parts of the
KST evolution equations, then, are given by:
\begin{eqnarray}
  \partial_t g_{ij} &\simeq& N^n\partial_n g_{ij}\ ; \label{eq:KSTg}
  \\
  \partial_t K_{ij} &\simeq& N^n\partial_n K_{ij} 
  -N
  \Bigl[
  (1+2\gamma_0) g^{cd} \delta^n_{\phantom{n}(i} \delta^b_{\phantom{b}j)}
  \nonumber\\ &&
  -(1+ \gamma_2) g^{nd} \delta^b_{\phantom{b}(i} \delta^c_{\phantom{c}j)} 
  -(1- \gamma_2) g^{bc} \delta^n_{\phantom{n}(i} \delta^d_{\phantom{d}j)} 
  \nonumber\\ &&
  +g^{nb} \delta^c_{\phantom{c}i} \delta^d_{\phantom{d}j}
  +2\gamma_1 g{}^{n[b}g{}^{d]c}g{}_{ij}
  \Bigr]
  \partial_n D_{bcd}\ ; \label{eq:KSTK}
  \\
  \partial_t D_{kij} &\simeq& N^n\partial_n D_{kij}
  -N
  \Bigl[
  \delta^n_{\phantom{n}k} \delta^b_{\phantom{k}i} \delta^c_{\phantom{j}j} 
  -\frac{1}{2}\, \gamma_3 g^{nb} g_{k(i}\delta^c_{\phantom{c}j)}
  \nonumber\\ &&
  -\frac{1}{2}\, \gamma_4 g{}^{nb}g{}_{ij}\delta{}^c{}_{k}
  +\frac{1}{2}\, \gamma_3 g{}^{bc}g{}_{k(i}\delta{}^n{}_{j)}
  \nonumber\\ &&
  +\frac{1}{2}\, \gamma_4 g{}^{bc}g{}_{ij}\delta{}^n{}_{k}
  \Bigr]
  \partial_n K_{bc}\ .\label{eq:KSTd}
\end{eqnarray}
Here, the symbol $\simeq$ indicates that terms algebraic in the fields
(that is, nonprincipal terms) are not shown explicitly.  The lapse
function $N$ is taken to be
\begin{equation}
  N \equiv g^{\gamma_0} e^Q\ ,
\end{equation}
and both the lapse density function $Q$ and the shift $N^i$ are
assumed to be specified functions of the coordinates, rather than
independent dynamical fields.  Since each of the Mexico City tests
involves reproducing either a known analytic solution of the Einstein
equations or a small perturbation about a known solution, for all
tests reported here we set the lapse density $Q$ and the shift $N^i$
from the appropriate analytic solution.  We choose one set of the KST
parameters for all the tests here: $\gamma_0=0.5$;
$\gamma_1=-0.21232$; $\gamma_2=-0.00787402$; $\gamma_3=-1.61994$;
$\gamma_4=-0.69885$.  These values were chosen because they make the
KST system symmetric hyperbolic and coincide with a set preferred by
Owen~\cite{Owen2006} in his extension of the KST system.

To evaluate errors it is useful to look at constraint quantities. As
mentioned above, the KST system has additional constraints,
Eqs.~(\ref{eq:ThreeIndexConstraint})
and~(\ref{eq:FourIndexConstraint}), besides the usual Hamiltonian
constraint $C$ and momentum constraint $C_i$.  To ensure that we are
satisfying all the constraints, we monitor a single quantity
$\mathcal{C}$ that is zero if and only if all of the constraints
vanish:
\begin{equation}\label{eq:C}
  \mathcal{C} \equiv \sqrt{C^2 + \left(C_i\right)^2 + 
    \left(C_{kij}\right)^2 + \left(C_{lkij}\right)^2}\ ,
\end{equation}
where an object is squared using the evolved spatial metric: for
example, $(C_i)^2 = g^{ij}C_i C_j$.

Likewise, when evaluating differences from analytically known
solutions, we define an overall error quantity that includes the
errors in all evolved variables $g_{ij}$, $K_{ij}$, and
$D_{kij}$. Taking $\delta g_{ij} \equiv g_{ij}^{\mathrm{analytic}} -
g_{ij}^{\mathrm{evolved}}$, and similarly for other fundamental
fields, this overall error quantity is given by
\begin{eqnarray}\label{eq:deltaU}
  \delta\, \mathcal{U} &\equiv& \sqrt{\left(\delta g_{ij}\right)^2 
    + \left(\delta K_{ij}\right)^2 + \left(\delta D_{kij}\right)^2}\ .
\end{eqnarray}
Notice that $\delta\, \mathcal{U}$ vanishes if and only if all evolved
variables match the known solution.

For all error quantities $\mathcal{Q}$ we display $L_2$ norms:
\begin{equation}
  \left\| \mathcal{Q} \right\|_2
  \equiv
  \sqrt{ 
    \frac{1}{\mathrm{Vol}}\, \int\, \mathcal{Q}^2\, \sqrt{|g|}\, d^3x 
  }\ ,
\end{equation}
where $\mathrm{Vol}=\int \sqrt{|g|} d^3x$ is the volume of the domain.
These norms are computed after each time step over the current
$t=\mbox{const.}$ hypersurface.  We refer to $\|\mathcal{C}\|_2$ as
the constraint energy, and $\|\delta\,\mathcal{U}\|_2$ as the error
energy.

The error quantities $\|\delta\,\mathcal{U}\|_2$ and
$\|\mathcal{C}\|_2$ scale with the absolute magnitude of the
fundamental fields and their derivatives, so it can be difficult to
judge the significance of these error measures without knowing the
overall scale of the variables in the problem.  For this reason, we
sometimes plot the \emph{normalized} error energy
$\|\delta\,\mathcal{U}\|_2/\|\mathcal{U}\|_2$ and the
\emph{normalized} constraint energy $\|\mathcal{C}\|_2/\|\partial\,
\mathcal{U}\|_2$, where the normalization factors are defined by
\begin{eqnarray}
  \mathcal{U} &\equiv& 
  \sqrt{\left(g_{ij}\right)^2 + \left(K_{ij}\right)^2 
    + \left(D_{kij}\right)^2
  }\ , \\
  \partial\, \mathcal{U} &\equiv& 
  \sqrt{\left(\partial_i g_{jk}\right)^2 
    + \left(\partial_i K_{jk}\right)^2 
    + \left(\partial_i D_{jkl}\right)^2
  }\ .
\end{eqnarray}
Note that $\|\delta\,\mathcal{U}\|_2/\|\mathcal{U}\|_2$ and
$\|\mathcal{C}\|_2/\|\partial\, \mathcal{U}\|_2$ become of order unity
when errors completely dominate the numerical solution.  We display
normalized error quantities only for tests involving the Gowdy
spacetimes (Section~\ref{sec:GowdySpacetime}), in which the
fundamental variables vary exponentially in time.  All other tests
presented here involve perturbations of Minkowski spacetime, in which
case the quantity $\|\partial\, \mathcal{U}\|_2$ is of order the size
of the perturbation and is therefore inappropriate to use as a
normalization factor.  However, for perturbations of Minkowski
spacetime, the overall scale is of order unity so it suffices to
display the unnormalized quantities $\|\delta\,\mathcal{U}\|_2$ and
$\|\mathcal{C}\|_2$.

\subsection{Pseudospectral Methods}

All of our numerical computations are carried out using pseudospectral
methods; this is the first time the Mexico City tests have been
applied to a pseudospectral code.  A brief outline of our method is as
follows: Given a system of partial differential equations
\begin{equation}
  \partial_t u(\mathbf{x},t) = 
  \mathcal{F}[u(\mathbf{x},t),\partial_i u(\mathbf{x},t) ]\ ,
  \label{diffeq}
\end{equation}
where $u$ is a collection of dynamical fields, the solution
$u(\mathbf{x},t)$ is expressed as a time-dependent linear combination
of $N$ spatial basis functions $\phi_k(\mathbf{x})$:
\begin{equation}
  u(\mathbf{x},t) = \sum_{k=0}^{N-1}\tilde{u}_k(t)\phi_k(\mathbf{x})\ .
  \label{decom}
\end{equation}
Associated with the basis functions is a set of $N_c$ collocation
points $\mathbf{x}_i$.  Given spectral coefficients $\tilde u_k(t)$,
the function values at the collocation points $u(\mathbf{x}_i,t)$ are
computed using Eq.~(\ref{decom}).  Conversely, the spectral
coefficients are obtained by the inverse transform
\begin{equation}
  \tilde{u}_k(t) =\sum_{i=0}^{N_c-1} w_i u(\mathbf{x}_i,t) 
  \phi_k(\mathbf{x}_i)\ , 
  \label{invdecom}
\end{equation}
where $w_i$ are weights specific to the choice of basis functions and
collocation points.  Thus it is straightforward to transform between
the spectral coefficients $\tilde{u}_k(t)$ and the function values at
the collocation points $u(\mathbf{x}_i,t)$.

To solve the differential equations, we evaluate spatial derivatives
analytically using the known derivatives of the basis functions,
\begin{equation}
  \partial_i u(\mathbf{x},t) 
  = \sum_{k=0}^{N-1}\tilde{u}_k(t)\partial_i\phi_k(\mathbf{x})\ ,
  \label{decomderiv}
\end{equation}
and we evaluate nonlinear terms using the values of
$u(\mathbf{x}_i,t)$ at the collocation points.  Thus we can write the
partial differential equation, Eq.~(\ref{diffeq}), as a set of
\emph{ordinary} differential equations for the function values at the
collocation points,
\begin{equation}
  \partial_t u(\mathbf{x}_i,t) = \mathcal{G}_i [u(\mathbf{x}_j,t)]\ ,
  \label{odiffeq}
\end{equation}
where $\mathcal{G}_i$ depends on $u(\mathbf{x}_j,t)$ for all $j$.  We
then integrate this system of ordinary differential equations in time,
using (for example) a fourth-order Runge-Kutta algorithm.

Because the tests discussed here are periodic in all spatial
dimensions, we use Fourier basis functions.  If we choose a
computational domain extending from $-1/2$ to $1/2$ in each of the
$x$-, $y$-, and $z$-directions, then each variable $u$ is decomposed
as
\begin{equation}
  u(x,y,z)
  =
  \sum_{k=0}^{N_x-1}\,
  \sum_{l=0}^{N_y-1}\,
  \sum_{m=0}^{N_z-1}\,
  a_{klm} \phi_k(x) \phi_l(y) \phi_m(z),
  \label{eq:FourierSeries}
\end{equation}
where
\begin{equation}
  \phi_k(x) = \left\{
    \begin{array}{ll}
      1                 & k=0\ ;                      \\
      \sin[\pi x (k+1)] & k>0\ \ (k \mathrm{~odd})\ ; \\
      \cos(\pi x k)     & k>0\ \ (k \mathrm{~even})\ .
    \end{array}\right.
  \label{eq:FourierBasisFns}
\end{equation}

For smooth solutions, the spectral approximation Eq.~(\ref{decom})
converges exponentially (error $\sim e^{-\lambda N}$ for some
$\lambda>0$ which depends on the solution).  This is much faster than
the polynomial convergence (error $\sim 1/N^p$) obtained using
$p$th-order finite-differencing.  As a result, we run our tests at
coarser resolutions than those recommended in
Ref.~\cite{Alcubierre2004} for finite-difference codes---typically we
use $N_i=9$, 15, 21, 27, and 33 collocation points in the relevant
directions.  From Eqs.~(\ref{eq:FourierSeries})
and~(\ref{eq:FourierBasisFns}) we see that if we choose $N_x$, $N_y$,
or $N_z$ to be an even integer, the highest frequency component in our
expansion will have a sine term but no matching cosine term.
Consequently the spatial derivative of this highest frequency
component will not be represented by our basis functions, causing a
numerical instability.  Therefore we choose $N_x$, $N_y$, and $N_z$ to
be odd.

Because spectral methods so greatly reduce spatial discretization
errors, time-stepping error is often dominant.  In order to make the
time stepping and the spatial discretization errors comparable in
these tests, we use fourth-order Runge-Kutta ODE integration.  The
time-step sizes are chosen in an effort to use step sizes comparable
to those used to test finite difference methods in
Ref.~\cite{Alcubierre2004}, while also ensuring that time-step errors
do not dominate over our spatial truncation errors.  We use $\Delta t
= \Delta x/20$ in the first test, and $\Delta t = \Delta x /40$ in all
others, except where explicitly noted.  Here, $\Delta x$ is the
minimum distance between collocation points.


\section{Random Initial Data on Flat Space}
\label{sec:RandomInitialData}

Perhaps the simplest test of a numerical relativity code is evolving
standard Minkowski spacetime on a three-torus, $T^3$.  However, this
test is \emph{too} simple because all fundamental fields are spatially
constant and most are identically zero, and hence most numerical
methods will reproduce the correct solution exactly.  This test can be
made more discriminating by adding a small amount of random noise to
the initial data; the noise is intended to simulate the effect of
finite numerical precision.  A different random number is added to
each component of each evolved variable, at each point in the
domain. These random numbers are chosen to lie between $-10^{-10}$ and
$10^{-10}$ so that the system remains in the linear regime.  If these
small perturbations to a simple spacetime grow unstably, it is likely
that the inevitable errors (e.g., discretization error or even
numerical roundoff error) that arise in any more complicated
simulation will also grow unstably.  For this test we vary the
resolution in the $x$-dimension, and we fix the resolution to three
collocation points in each of the $y$- and $z$-dimensions.

\begin{figure}
  \begin{center}
    \includegraphics[width=\linewidth, clip=false, trim= 0 0 0 0]%
    {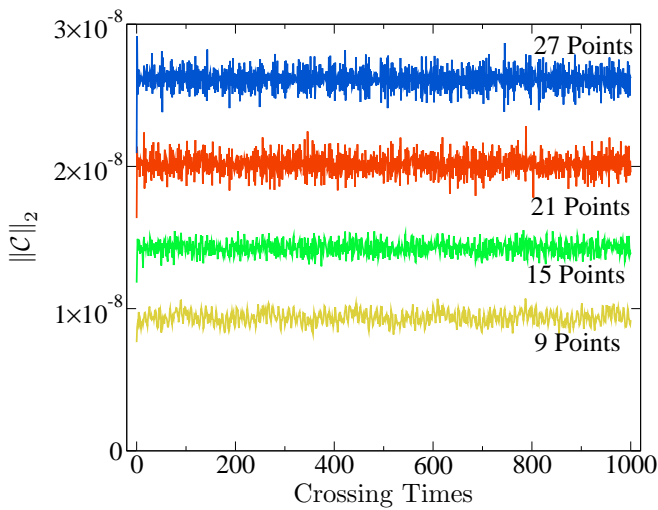}
  \caption{\textbf{Constraints for Minkowski with Random Noise.}
    Higher resolutions are expected to have larger constraints because
    more closely spaced points result in larger derivatives.  The 
    constraints do not grow in time.
    \label{fig:RobustStabilityCE}}
\end{center}
\end{figure}

If the perturbations in the fields are chosen to be of size $\epsilon$
independent of resolution, then the perturbation in the $n$th spatial
derivatives of these fields will be $\sim \epsilon (\Delta x)^{-n}$,
where $\Delta x$ is some measure of the distance between neighboring
points.  This means that error quantities involving derivatives (such
as constraints) will be \emph{larger} for finer resolutions.%
\footnote{The Mexico City tests collaboration~\cite{Alcubierre2004}
  intended their Hamiltonian constraint errors to be
  resolution-independent, so they chose the size of the perturbation
  $\epsilon$ to be resolution-\emph{dependent}, $\epsilon \sim (\Delta
  x)^2$, which is the appropriate scaling for the
  second-order-in-space formulations of Einstein's equations they use.
  However, for the first-order-in-space formulation we use, the
  Hamiltonian constraint is computed using first derivatives of
  $D_{kij}$ rather than second derivatives of $g_{ij}$, so the
  constraint will vary as $(\Delta x)^{-1}$.  Note also that the
  $\epsilon \sim (\Delta x)^2$ scaling does not make the momentum
  constraint independent of resolution, as it depends on first
  derivatives of the fields.  We simply choose $\epsilon$ to be
  independent of resolution.}  This behavior is seen in the plot of
the constraint energy in Fig.~\ref{fig:RobustStabilityCE}.

The purpose of this test is to establish that small constraint
violations around flat space do not grow, and the KST system clearly
passes this test.  Whether or not constraint violations decay will
depend on the evolution system and the numerical method.  For example,
artificial dissipation in the numerical method might cause all
variations to decay, including constraint violations.  Furthermore, if
the evolution system contains constraint damping in some form, then
the constraints should decay.  Indeed, Owen has extended the KST
system to include constraint damping~\cite{Owen2006}; running the same
test, he observes exponential decay in the constraint quantities.  The
flat constraint violations observed in
Fig.~\ref{fig:RobustStabilityCE} indicate that the KST system with our
parameter choice does not damp constraints and that the spectral
method has insignificant artificial dissipation.

\begin{figure}
  \begin{center}
    \includegraphics[width=\linewidth, clip=false, trim= 0 0 0 0]%
    {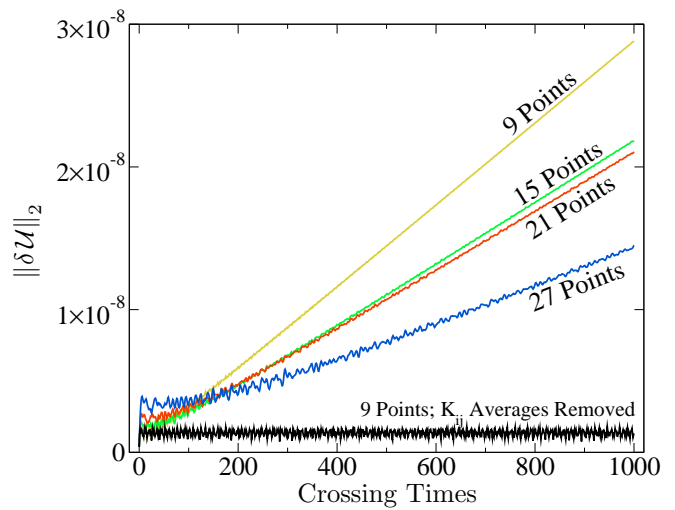}
  \caption{\textbf{Error Energy for Minkowski with Random Noise.}  
    The linear increase in time is due to a nonzero average in the 
    random noise added to $K_{ij}$.  This average approaches zero as 
    resolution is increased, since there are more points over which to
    average.  The flat line shows the evolution when the average 
    value of $K_{ij}$ is set to zero in the initial data.
    \label{fig:RobustStabilityEE}}
\end{center}
\end{figure}

In Fig.~\ref{fig:RobustStabilityEE} we see a linear growth of the
error energy $\|\delta\, \mathcal{U}\|$ for this test.  We find that
the growth is caused solely by contributions from the metric $g_{ij}$;
the average values of $K_{ij}$ and $D_{kij}$ remain constant in time.
We can understand this as follows: The average value of $K_{ij}$ is
determined by the random initial data and will in general be nonzero.
The time derivative of $K_{ij}$, to first order in the amplitude of
perturbations around flat space, involves only spatial derivatives of
$D_{kij}$.  These derivatives have zero average (up to roundoff errors
$\sim 10^{-16}$), because the constant term in the Fourier expansion
Eq.~(\ref{eq:FourierBasisFns}) is removed by differentiation, and
therefore the average of $K_{ij}$ will be constant in time.  The time
derivative of $g_{ij}$ involves a term proportional to $K_{ij}$.
Because the average of $K_{ij}$ is constant in time and nonzero, the
value of $g_{ij}$ will therefore drift linearly in time.  The average
of $K_{ij}$ is smaller for higher resolutions---because one averages
over more random numbers---which means that the growth rate of
$g_{ij}$ should decrease with increasing resolution.  Indeed, this is
what we observe in Fig.~\ref{fig:RobustStabilityEE}.

We can verify that the nonzero average of $K_{ij}$ is the only source
of growth in $g_{ij}$ by manually removing the average value of
$K_{ij}$. We expect this will leave the norms of the components of
$g_{ij}$ approximately constant in time.  This is accomplished by
setting the $k=0$ spectral coefficients of all components of $K_{ij}$
to zero in the initial data, after all the random numbers have been
added.  The flat line in Fig.~\ref{fig:RobustStabilityEE} shows the
result, indicating that the average offset in $K_{ij}$ is the only
source of growth in the evolved variables of the KST system for this
test.


\section{Linear Plane Wave}
\label{sec:LinearWave}

If the ultimate goal of simulating binary black hole mergers is to
predict the gravitational-radiation waveforms for observations, an
evolution system must at least be capable of evolving a simple linear
plane wave through flat spacetime.  The form suggested for the Mexico
City tests in Ref.~\cite{Alcubierre2004} is
\begin{equation}
  ds^2 = -dt^2 + dx^2 + (1+b) dy^2 + (1-b) dz^2\ ,
  \label{eq:LinearWaveMetric}
\end{equation}
where
\begin{equation}
  b = b(x,t) = A \sin \left[ 2\pi (x-t) \right]\ .
  \label{eq:LinearWaveMetricFunction}
\end{equation}
This metric satisfies Einstein's equations only to linear order in the
wave amplitude $A$, so if the fully nonlinear numerical solution is
compared to this approximate solution, there will be deviations of
order $A^2$ that arise from our choice of ``analytic solution'' rather
than from numerical errors.  The amplitude $A$ for the Mexico City
tests is chosen to be $10^{-8}$ so that such deviations in the metric
components $g_{ij}$ are below machine precision. However, we still
observe $\mathcal{O}(A^2)$ deviations in the variables $K_{ij}$ and
$D_{kij}$ (which have values of order $A$), even with $A=10^{-8}$,
because the \emph{relative} error is well above machine precision.

\subsection{One-Dimensional Sinusoid}
\label{subsec:LinearWave1D}

The sinusoidal waveform chosen in
Eq.~(\ref{eq:LinearWaveMetricFunction}) is only a weak test for
pseudospectral methods because the Fourier basis functions defined in
Eqs.~(\ref{eq:FourierSeries}) and~(\ref{eq:FourierBasisFns}) exactly
resolve Eq.~(\ref{eq:LinearWaveMetricFunction}) at all times using
only three basis functions; the only truncation errors are those
associated with time discretization.  Therefore, as a more challenging
test, in Section~\ref{subsec:LinearWaveGaussian} we repeat the plane
wave evolution using a Gaussian-shaped wave.  It is nevertheless
instructive to evolve the sinusoid and study the resulting time
discretization errors.  Since the dynamics involve no change in
amplitude, but a change in phase, we expect the errors to be primarily
phase errors, for reasonably small time steps.

\begin{figure}
  \begin{center}
    \includegraphics[width=\linewidth, clip=false, trim= 0 0 0 0]%
    {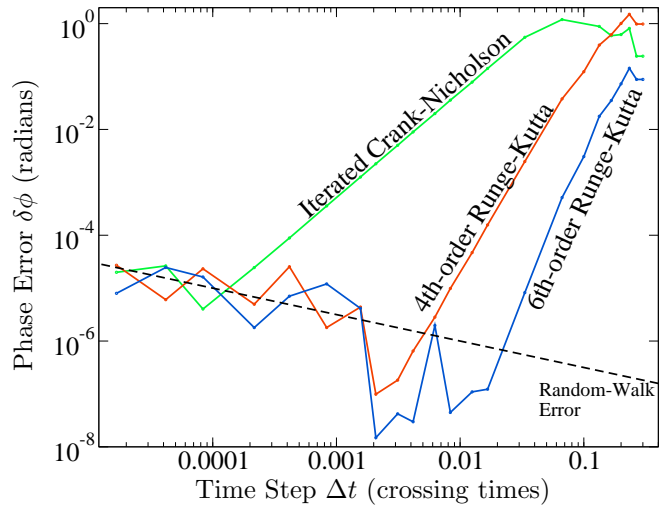}
  \caption{\textbf{Phase Error for 1-D Sinusoidal Linear Wave.}  
    Phase error at $t=25$ crossing times for various time steps, and 
    several time-stepping algorithms.  These tests were all run with 
    three points in the \mbox{$x$-direction}.  The dashed line 
    indicates the expected accuracy limit due to roundoff error, cf. 
    Eq.~(\ref{eq:PhaseError}).
    \label{fig:LinearWavePhaseError}}
\end{center}
\end{figure}

This loss of temporal accuracy is particularly relevant in efforts to
simulate sources for gravitational wave observations, as the search
for signals involves matching expected waveforms against observations.
If there is significant error in the phase of the expected waveform,
the overlap will be poor and detection will be more difficult.
Although a constant overall scaling error in frequency---like the one
found in this linear problem---could still result in detection, more
complex situations would likely give rise to more complicated errors.
The straightforward way to handle this problem is to minimize all
time-stepping error.

In Fig.~\ref{fig:LinearWavePhaseError} we show the convergence of the
phase error in the evolution of the sinusoidal linear wave.  The
solution is fully resolved on a $3\times 1\times 1$ grid.  We keep
this grid fixed, and decrease the size of the time step.  Assuming
that the only error is some phase error $\delta \phi$, the evolved
$g_{zz}$ will be given by
\begin{equation}
  g_{zz} = 1 - 10^{-8} \sin \left[ 2\pi(x-t) + \delta \phi \right]\ .
\end{equation}
At integer multiples of the light crossing time for our computational
domain, this can be written as
\begin{equation}
  g_{zz}=
  1 - A \left[ \cos \delta \phi \sin \left( 2\pi x \right) 
    + \sin \delta \phi \cos \left(2\pi x \right) \right]\ .
  \label{eq:gzzError}
\end{equation}
That is, we can find the phase error easily from the $k=1$ sine and
cosine components of $g_{zz}$ (which happen to be easily accessible in
our code).

For intermediate time-step sizes, we can see convergence toward zero
phase error with decreasing time step.  As expected, we observe
second-order convergence for Iterated Crank-Nicholson stepping, and
fourth- and sixth-order for the appropriate higher-order Runge-Kutta
algorithms.  At very small time-step sizes, a new effect is seen,
causing the phase error to increase with decreasing time step.  This
effect can be understood as machine roundoff error accumulating via a
random walk process.

Suppose we have a variable $\mathcal{V}(t)$ that is evolved by adding
the small changes needed to update its value at each time step.  Each
such operation will introduce a fractional error $\chi(t)$ caused by
the finite machine precision.  We assume that the standard time-step
size is $\Delta t$, and that there are $n$ intermediate operations in
each time step.  After an evolution through time $T$, the total error
added in this way will be
\begin{equation}
  \delta \mathcal{V}
  =
  \sum_{j=0}^{n\, T/\Delta t}\, \mathcal{V}(t_j)\, \chi(t_j)\ .
\end{equation}
To avoid tracking each individual error contribution, we treat $\chi$
as a random variable taking values in some range, with some
probability distribution.

We estimate the accumulated error due to finite machine precision by
taking suitable averages over an ensemble of random $\chi(t)$ and over
a time interval $T$.  If there were no asymmetry between positive and
negative values of $\chi(t)$, this accumulated error would be zero.
Of course, we expect almost never to see this case: the most likely
outcome is an accumulated error comparable to the dispersion:
\begin{equation}
  \left| \delta \mathcal{V} \right|
  \sim
  \sqrt{\overline{(\delta \mathcal{V})^2}}
  \sim
  \sqrt{\sum_{j,k}\, \mathcal{V}(t_j)\, \mathcal{V}(t_k)\, 
    \overline{\chi(t_j) \chi(t_k)}}\ ,
\end{equation}
where the overbar indicates the average over the ensemble of random
errors $\chi(t)$.  We can simplify this expression by assuming that
$\chi(t)$ has no correlations between time steps, and further assuming
that the probability distribution is constant in time and uniform,
taking values in the range $[-\epsilon, \epsilon]$, where $\epsilon$
is the machine precision.  This means that $\overline{\chi(t_j)
  \chi(t_k)} = \delta_{jk} \epsilon^2/3$.  Finally, we approximate the
discrete time sum as an integral, and obtain
\begin{equation}
  \left| \delta \mathcal{V} \right|
  \sim
  \epsilon\, \sqrt{\frac{n}{3\Delta t}\, 
    \int_{t_1}^{t_2}\, \mathcal{V}(t)^2\, dt}\ .
  \label{eq:Error}
\end{equation}

We can test this formula by observing its effects in the case of phase
error for the linear wave.  Here, the only nontrivial evolved variable
is $\mathcal{V}= g_{zz}$, which is very nearly 1; so the integral in
Eq.~(\ref{eq:Error}) becomes simply the evolution time $T$, which has
the value $25$ for the results plotted in
Fig.~\ref{fig:LinearWavePhaseError}.  If phase errors dominate,
$\delta g_{xx}\sim A\sin\delta\phi$, so we have
\begin{equation}
  |\delta \phi|
  \sim  \frac{\epsilon}{A}
  \sqrt{\frac{25\,n} {3\, \Delta t}}\ \sim
  \frac{10^{-7}}{\sqrt{\Delta t}}
  \left(\frac{\epsilon}{10^{-16}}\right)\left(\frac{10^{-8}}{A}\right)
  \ ,
  \label{eq:PhaseError}
\end{equation}
where $n$ is assumed to be of order $10$.  This expression is plotted
as the dashed line in Fig.~\ref{fig:LinearWavePhaseError},
demonstrating that the addition of machine-precision errors causes the
departure from the standard second-, fourth-, and sixth-order
convergence we observed.  From Eq.~(\ref{eq:PhaseError}) we see that
$|\delta\phi|$ is proportional to the ratio $\epsilon/A$; thus
$|\delta\phi|$ is so large in this case because the wave amplitude is
so small, $A=10^{-8}$.

The phase error is only so clearly visible in these evolutions because
the full solution is described precisely at each moment by the first
three basis functions.  This means that discretization error due to
spatial differentiation is essentially at the level of machine
precision.  Indeed, using more than three points actually degrades the
quality of these one-dimensional sinusoid evolutions.  Power in
higher-order basis functions can only be error, and hence will
necessarily do worse than the low-resolution case.  We omit plots of
the error energy and constraints in the higher-resolution cases, as
they are very nearly the same as those of the more complicated
two-dimensional evolutions discussed in
Section~\ref{subsec:LinearWave2D}.

\subsection{One-Dimensional Gaussian}
\label{subsec:LinearWaveGaussian}

As a more challenging test of pseudospectral methods, we repeat the
one-dimensional linear wave test using a periodic Gaussian-shaped
wave:
\begin{equation}
  b = A\, \sum_{j=-\infty}^{\infty}\, 
  \exp \left[ -\frac{\left( x - t + j \right)^2} {2\, w^2} \right]\ ,
\end{equation}
with $A=10^{-8}$.  The summation over $j$ ensures that the function is
truly periodic at all times.  In practice, $j$ need only range over a
few, depending on the width of the Gaussian.  The width chosen here is
$w=0.05$ to ensure that features are not too sharp, while still
presenting a nontrivial challenge to spectral differentiation.

\begin{figure}
  \begin{center}
    \includegraphics[width=\linewidth, clip=false, trim= 0 0 0 0]%
    {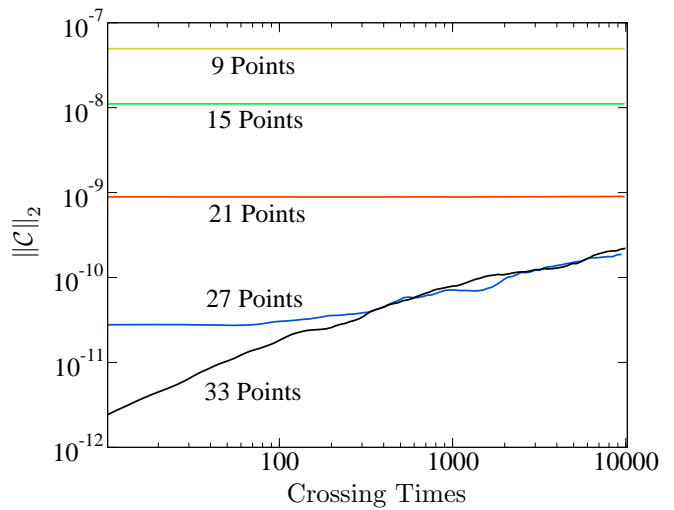}
  \caption{\textbf{Constraints for 1-D Gaussian Linear Wave.}  
    Here we see the exponential convergence of the constraints with
    higher spatial resolution.  At late times, the higher resolutions
    grow sublinearly in time, probably because of accumulated machine
    roundoff error. 
    \label{fig:LinearWaveGaussianCE}}
\end{center}
\end{figure}

\begin{figure}
  \begin{center}
    \includegraphics[width=\linewidth, clip=false, trim= 0 0 0 0]%
    {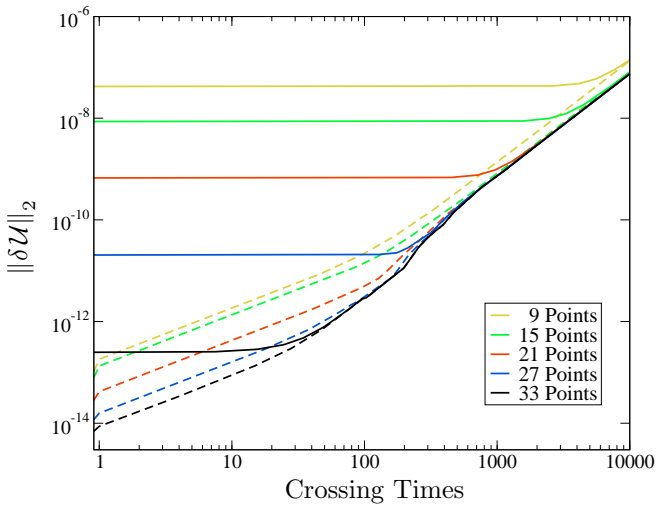}
  \caption{\textbf{Error Energy for 1-D Gaussian Linear Wave.} 
    The solid lines show the error energy at 1/2 crossing times, with 
    clearly visible exponential convergence at early times.  The 
    dashed lines show the error energy at integer crossing times for 
    the same resolutions.  The smallness of the error energy at early 
    times demonstrates the low dispersion of the numerical method, as 
    explained in the main text.  At later times, the error is 
    dominated by the quadratic growth explained in the text.
    \label{fig:LinearWaveGaussianEE}}
\end{center}
\end{figure}

We find behavior comparable to the sinusoidal case, although as
expected, more collocation points in the $x$-direction are needed to
resolve the solution spatially (but we still use only a single point
in each of the $y$- and $z$-dimensions).  Note the exponential
convergence of the constraints with increasing resolution in
Fig.~\ref{fig:LinearWaveGaussianCE}.  The constraint growth in the
highest resolution runs is slower than linear in time, and is probably
caused by the accumulation of errors due to finite machine precision
as discussed in Sec.~\ref{subsec:LinearWave1D}.

Figure~\ref{fig:LinearWaveGaussianEE} presents the error energy for
this run as the solid lines.  At early times $||\delta \mathcal{U}||$
decreases with resolution exponentially to zero, as one would expect.
At late times, however, $||\delta \mathcal{U}||$ converges toward a
parabola.  The amplitude of this parabola scales in proportion to
$A^2$.  In the rest of this section, we will first explain a subtlety
arising when computing $||\delta \mathcal{U}||$, followed by a
detailed explanation of why the terms $\mathcal{O}(A^2)$ manifest
themselves in parabolic behavior of $||\delta \mathcal{U}||$.

The comparison of the computed solution with the analytic solution is
performed at the collocation points.  By virtue of the transformation
Eqs.~(\ref{decom}) and (\ref{invdecom}), the errors are initially
exactly zero at the collocation points.  The spatial truncation error
is nonzero of course, even at the initial time; it manifests itself by
a deviation of the truncated series expansion from the analytic
solution \emph{between} collocation points.  During the evolution, a
linear wave will simply travel through the computational domain,
returning to the original position after each light-crossing time.
Since the spectral method has very small dispersion, the evolved shape
remains the same.  After each light-crossing time, therefore, the
evolved solution again agrees to very high accuracy with the initial
analytic solution \emph{at} the collocation points.  So, comparing the
evolution with the analytic solution at integer multiples of the
light-crossing time and at the collocation points will yield
differences much smaller than spatial truncation error.%
\footnote{This is also true when comparing at intervals of $1/N$ of a
  crossing time if the number of collocation points in the direction
  of the wave's travel is divisible by $N$.}  Therefore, a fair
comparison that includes the effects of spatial truncation error must
not be performed at integer light-crossing times.  These
considerations are evident from Fig.~\ref{fig:LinearWaveGaussianEE},
where the solid lines show the ``true'' $||\delta \mathcal{U}||$
observed with 1/2 light-crossing interval offset, which suffices
because the number of collocation points is always odd.  The
artificially small error energy observed at every complete
light-crossing interval is shown as dashed lines, confirming the
excellent low-dispersion property of our method.

At late times the differences between observation at full and 1/2
crossing times are swamped by the parabolic growth in $||\delta
\mathcal{U}||$.  Similar parabolic deviations of the evolution from
the solution of the linearized equations are observed for the other
two linear wave evolutions, the \mbox{1-D} and \mbox{2-D} sinusoids
(cf. Fig.~\ref{fig:LinearWaveEE}).  The growth in $\|\delta\,
\mathcal{U}\|$ appears almost entirely due to growth in the $k=0$ mode
of diagonal terms in $\delta g_{ij}$.  Using evolutions of waves with
different amplitudes and wavelengths, we have verified that this
growth is proportional to $A^2t^2/\lambda^2$, where $A$ is the
amplitude and $\lambda$ the wavelength of the disturbance. The
constant of proportionality depends directly on the KST parameter
$\gamma_1$ appearing in Eq.~(\ref{eq:KSTK}).  This parameter controls
the addition of a term $\gamma_1 N g_{ij} C$ to the ADM evolution
equation for $K_{ij}$.  The Hamiltonian constraint $C$ is roughly
constant in time, and varies as $A^2 / \lambda^2$.  Since the $k=0$
mode of $\gamma_1 N g_{ij} C$ is roughly $\gamma_1 \delta_{ij} C$, the
$k=0$ mode of $K_{ii}$ will grow linearly with time in proportion to
$\gamma_1 A^2 / \lambda^2$ for each $i$.  That, in turn, will cause
small quadratic growth in the $k=0$ mode of $g_{ii}$.  For the more
well-behaved cases (highest resolutions for the Gaussians; all cases
for the sinusoids) this model is an excellent fit for the observed
error energy.

\subsection{Two-Dimensional Linear Waves}
\label{subsec:LinearWave2D}

\begin{figure}
  \begin{center}
    \includegraphics[width=\linewidth, clip=false, trim= 0 0 0 0]%
    {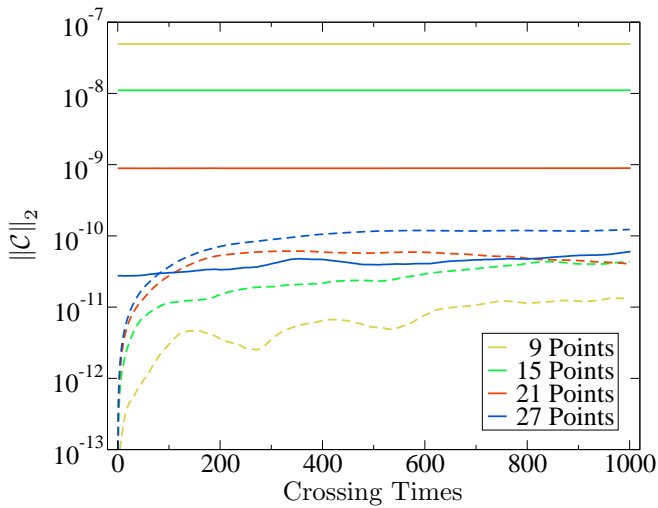}
  \caption{\textbf{Constraints for 2-D Linear Waves.}
    The solid lines represent the Gaussian wave, while the dashed 
    lines represent the sinusoidal wave.
    The sinusoid is fully resolved spatially with 3 points.  Going to 
    higher resolutions merely introduces spatial errors in the 
    unnecessary basis functions, which leads to an increase in the 
    constraints with resolution.
    \label{fig:LinearWaveCE}}
\end{center}
\end{figure}

\begin{figure}
  \begin{center}
    \includegraphics[width=\linewidth, clip=false, trim= 0 0 0 0]%
    {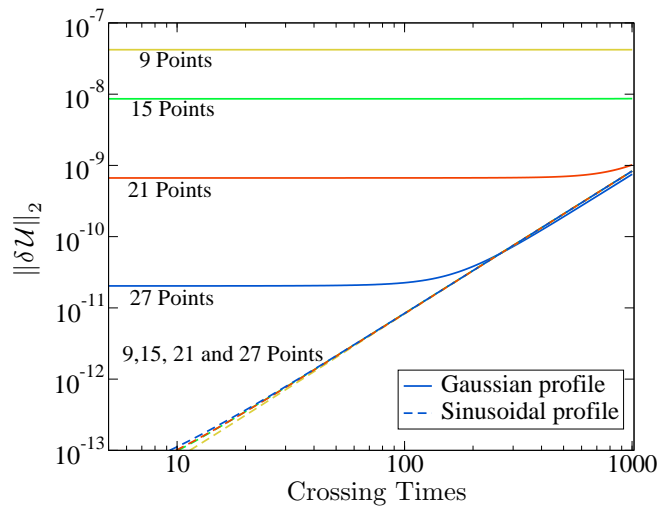}
  \caption{\textbf{Error Energy for 2-D Linear Waves.}  The solid 
    lines represent the Gaussian wave, while dashed lines represent 
    the sinusoidal wave, both observed at 1/2 crossing times. As in 
    the 1-D case (Fig.~\ref{fig:LinearWaveGaussianEE}), both sets of 
    evolutions converge to quadratic growth of the error caused by 
    the Hamiltonian constraint, explained in the text.
    \label{fig:LinearWaveEE}}
\end{center}
\end{figure}

The linear wave tests above may be modified by rotating the
coordinates by $\pi/4$ about the $z$-axis, which gives a plane wave
propagating along the \mbox{$x$-$y$} diagonal.  By increasing the size
of the domain by a factor of $\sqrt{2}$ in each direction, the rotated
solution can be made periodic while maintaining the same wavelength.
This converts the spacetime from essentially one-dimensional to
essentially two-dimensional.  The purpose of this test is to ensure
that the symmetry of the one-dimensional version does not hide sources
of error (although propagation along a diagonal retains some
symmetries).  For these tests we use a single collocation point in the
$z$-dimension, and we vary the (equal) number of collocation points in
the $x$- and $y$-dimensions.  We run these tests to $t=1000$---ten
times longer than is recommended by the Apples with Apples
collaboration---to better observe the stability properties.  As shown
in Fig.~\ref{fig:LinearWaveCE}, the constraints for the sinusoidal
wave increase with increasing $x$- and $y$-resolution (still using
only a single point in the $z$-direction).  The constraints for the
Gaussian are very nearly the same as in the one-dimensional case.
Again, the $A^2 t^2/\lambda^2$ growth of $\|\delta\,\mathcal{U}\|$ is
visible, as shown in Fig.~\ref{fig:LinearWaveEE}.

\section{Gauge Wave}
\label{sec:GaugeWave}

The next series of tests involves a simple but time-dependent gauge
transformation of Minkowski space, in the form of a plane wave.  The
metric used for this Mexico City test has the form
\begin{equation}
  ds^2 = -(1+a) dt^2 + (1+a) dx^2 + dy^2 + dz^2\ ,
  \label{eq:GaugeWaveMetric}
\end{equation}
\begin{equation}
  a = A \sin \left[ 2\pi (x-t) \right]\ .
  \label{eq:GaugeWaveMetricFunction}
\end{equation}
Two cases are considered: a low amplitude case $A=0.01$, and a high
amplitude case $A=0.1$.  This is the first test for which the
nonlinear terms in the equations play an important role.

For the linear plane wave test in Section~\ref{sec:LinearWave}, we
found that because we use a Fourier basis, we were able to fully
resolve the sinusoidal waveform using only three collocation points.
This is not true for the gauge-wave test, because in this case the
extrinsic curvature (one of our evolved variables) is not a simple
sinusoid. Instead, its only nonzero component is
\begin{equation}
  K_{xx} = - \pi\, \frac{A\, \cos \left[ 2\pi (x-t) \right]} 
  {\sqrt{1 + A\, \sin \left[ 2\pi (x-t) \right]}}\ .
\end{equation}

\subsection{One-Dimensional Gauge Wave}
\label{subsec:GaugeWave1D}

We ran the one-dimensional test described above using a single
collocation point in each of the $y$- and $z$-dimensions, and varying
the resolution in the $x$-dimension.  We find that for both $A=0.01$
and $A=0.1$ our evolution is stable and convergent.  Our error energy
and constraint violations show no signs of instability and are
strictly better than the filtered two-dimensional evolutions discussed
below.  We omit plots for this test because the two-dimensional test
is more challenging and more discriminating.

\subsection{Two-Dimensional Gauge Wave}
\label{subsec:GaugeWave2D}

\begin{figure}
  \begin{center}
    \includegraphics[width=\linewidth, clip=false, trim= 0 0 0 0]%
    {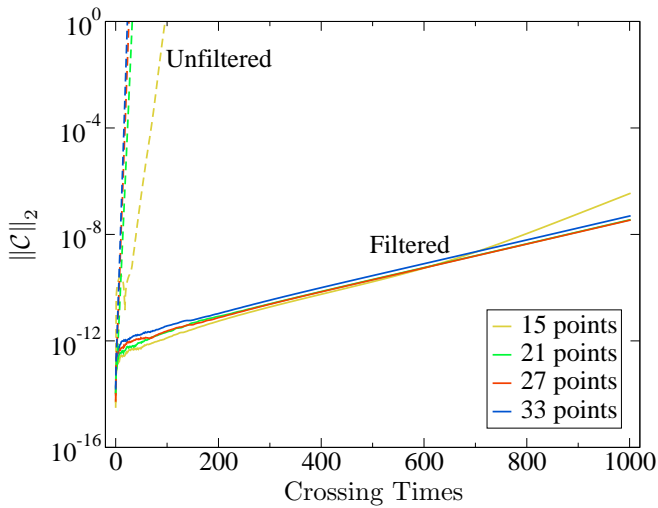}
  \caption{\textbf{Constraints for High-Amplitude 2-D Gauge Wave.}  
    Dashed lines indicate the unfiltered behavior; solid lines 
    indicate the filtered behavior.  Note that, despite an effective 
    loss of resolution, filtering greatly improves the stability of 
    the evolution.
    \label{fig:GaugeWave2DCE}}
\end{center}
\end{figure}

\begin{figure}
  \begin{center}
    \includegraphics[width=\linewidth, clip=false, trim= 0 0 0 0]%
    {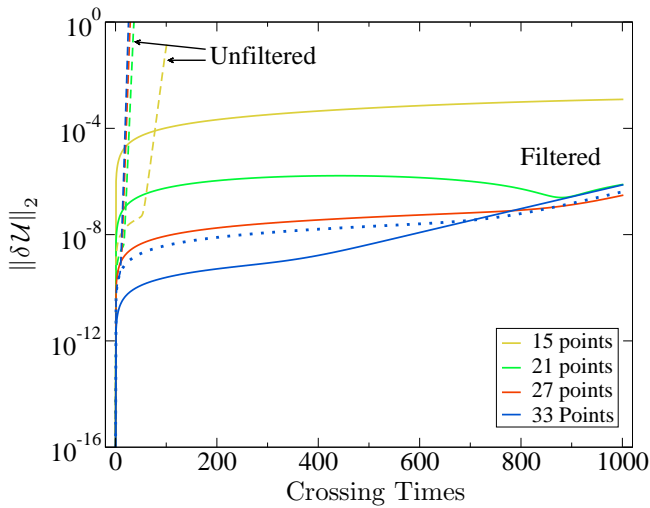}
  \caption{\textbf{Error Energy for High-Amplitude 2-D Gauge Wave.}  
    As in Fig.~\ref{fig:GaugeWave2DCE}, 
    dashed lines are unfiltered, and solid lines are filtered. 
    The growth of the filtered error energy is exponential in time.  
    For the highest resolution the time step was cut in half 
    ($dt=dx/80$) to reduce time-discretization error to the same level
    as spatial discretization error.  The dotted line shows the same 
    evolution with time step $dt=dx/40$, which is dominated by time 
    discretization error.
    \label{fig:GaugeWave2DEE}}
\end{center}
\end{figure}

A simple rotation of coordinates about the $z$-axis makes the wave
described by Eqs.~(\ref{eq:GaugeWaveMetric})
and~(\ref{eq:GaugeWaveMetricFunction}) propagate along the $x$-$y$
diagonal, as in the case of the linear wave.  We use an equal number
of collocation points in the $x$- and $y$-dimensions, and a single
collocation point in $z$.

As for the one-dimensional gauge wave test, we run at two different
amplitudes: $A=0.01$ and $A=0.1$. For low amplitude, $A=0.01$, our
evolution of the 2-D gauge wave is stable and convergent.  Again, we
omit plots, as our results are strictly better than for the more
interesting high-amplitude case.

For high amplitude, $A=0.1$, we find an exponentially growing
nonconvergent numerical instability, as seen in the curves labeled
``unfiltered'' in Figs.~\ref{fig:GaugeWave2DCE}
and~\ref{fig:GaugeWave2DEE}.  This instability does not appear for the
low-amplitude case, nor does it appear for either amplitude in the
one-dimensional gauge wave test.

The instability appears to be associated with aliasing caused by
quadratic nonlinearities in the evolution equations; this is a
well-known phenomenon that often occurs when applying spectral methods
to nonlinear equations~\cite{Boyd1989}.  The basic mechanism for the
instability can be understood by considering a truncated spectral
expansion for some variable $u(x)$ in terms of $N$ basis functions
$\phi_k(x)$:
\begin{equation}
  \label{eq:OneDSpectralExpansion}
  u(x) = \sum_{k=0}^{N-1} u_k \phi_k(x)\ .
\end{equation}
The correct spectral expansion of the expression $u(x)^2$ can be
obtained by squaring Eq.~(\ref{eq:OneDSpectralExpansion}); for most
basis functions---including the Fourier series of
Eq.~(\ref{eq:FourierBasisFns})---this yields a sum over a total of
$2N$, and not just $N$, basis functions.  But we keep only $N$ basis
functions (not $2N$) in our expansion, so the $k\ge N$ coefficients of
the product must be eliminated. Ideally, these $k\ge N$ coefficients
should be simply discarded and the $k<N$ coefficients should remain
untouched. But it turns out that for the pseudospectral method of
evaluating nonlinear terms (i.e., Fourier transform to obtain values
at spatial collocation points, square these values, then Fourier
transform back to spectral space), the power in the extra $k\ge N$
coefficients of the product does not disappear, but instead appears as
excess power in some of the $k<N$ coefficients (``aliasing'').
Repeating this process on each time step builds up this excess power
and produces the instability.

Fortunately, there is a well-known remedy for instabilities caused by
aliasing in nonlinear terms: suppose that the upper half (i.e., those
with $k\ge N/2$) of the coefficients $u_k$ in
Eq.~(\ref{eq:OneDSpectralExpansion}) were all zero.  Then the spectral
expansion of $u(x)^2$ will have zeroes in all its $k\ge N$
coefficients, so there is no aliasing, and hence no instability.
Therefore, we ensure that all coefficients with $k\ge
k_{\mathrm{cut}}$ are zero by removing those coefficients from the
initial data and from the right-hand side of the evolution equations.
It turns out (see, for example, Chapter~11.5 of Ref.~\cite{Boyd1989})
that for a quadratic nonlinearity, it is sufficient to filter with
$k_{\mathrm{cut}}= 2N/3$ (and not $k_{\mathrm{cut}}=N/2$) to eliminate
aliasing.  As mentioned in
Section~\ref{sec:AdaptingTheMexicoCityTests}, the remaining number of
nonzero coefficients must be odd, which is ensured by reducing
$k_{\mathrm{cut}}$ by one if necessary.

The price we pay for stability via this filtering is that we must use
a factor of $1.5$ more spectral coefficients (and collocation points)
than without filtering in order to achieve the same level of spatial
discretization error.  Hence, we use more points for this test than
for the previous ones: $N_i=15$, 21, 27, and 33 points.  This leaves
the effective resolutions at $\tilde{N}_i=9$, 13, 17, and 21 points,
which are comparable to the resolutions we use in the unfiltered case.
We see from Figs.~\ref{fig:GaugeWave2DCE} and~\ref{fig:GaugeWave2DEE}
that filtering dramatically reduces the instability.  The initial
constraint violations in these runs, $\| \mathcal{C} \| \approx
10^{-12}$, are at the level of the finite machine precision, so
increasing the resolution causes \emph{increased}---not
decreased---constraint violations.

\subsection{Shifted Gauge Wave}
\label{subsec:ShiftedGaugeWave}

\begin{figure}
  \begin{center}
    \includegraphics[width=\linewidth, clip=false, trim= 0 0 0 0]%
    {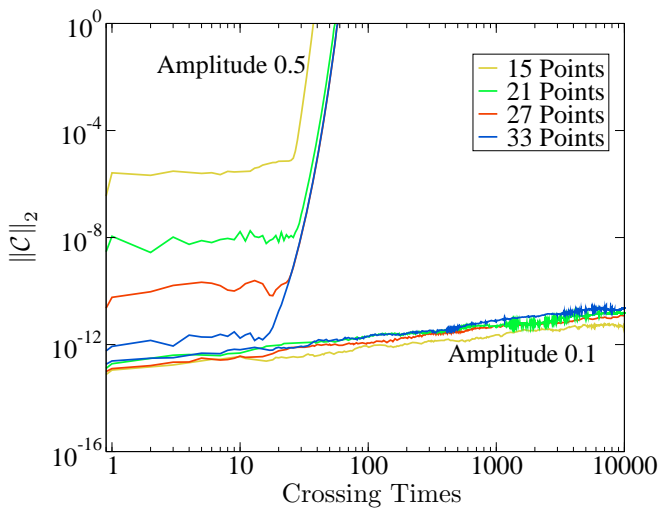}
  \caption{\textbf{Constraints for Shifted Gauge Wave.}
    Solid lines indicate $A=0.5$, and dashed lines indicate $A=0.1$.  
    For both amplitudes we filter out the top $1/3$ spectral 
    coefficients.
    \label{fig:ShiftedGaugeWaveCE}}
\end{center}
\end{figure}

\begin{figure}
  \begin{center}
    \includegraphics[width=\linewidth, clip=false, trim= 0 0 0 0]%
    {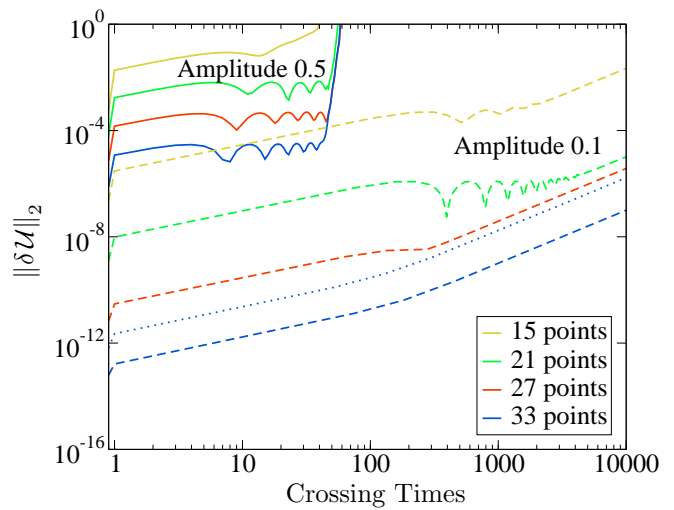}
\caption{\textbf{Error Energy for Shifted Gauge Wave.}  
Solid lines indicate $A=0.5$, while dashed
  lines indicate $A=0.1$.  The growth in the $A=0.1$ runs is roughly
  linear in time, accelerating to quadratic at later times.  The
  dotted line indicates the standard time step ($dt=dx/40$) with $33$
  points, which is dominated by temporal discretization error, while
  the blue dashed curve uses $dt=dx/80$.
  \label{fig:ShiftedGaugeWaveEE} }
\end{center}
\end{figure}

We also show the results of a new ``shifted gauge wave'' test
suggested for addition to the ``Apples with Apples''
suite~\cite{Babiuc2006}.  For this test we evolve flat space with the
usual Minkowski coordinates $(\hat{t},\hat{x},\hat{y},\hat{z})$
transformed to coordinates $(t,x,y,z)$ via
\begin{eqnarray}
  \hat{t}&=& t - \frac{A}{4\pi}\, \cos \left[ 2\pi(x-t) \right]\ , \\
  \hat{x}&=& x - \frac{A}{4\pi}\, \cos \left[ 2\pi(x-t) \right]\ , \\
  \hat{y}&=& y\ , \\
  \hat{z}&=& z\ .
\end{eqnarray}
This test includes the effects of a nonvanishing shift vector.  We use
the same computational domain and KST parameters as in the standard
gauge wave tests above.  The amplitude suggested in
Ref.~\cite{Babiuc2006} is $A=0.5$.  We also run simulations with
$A=0.1$.

At high amplitude, $A=0.5$, we see exponentially growing nonconvergent
instabilities.  Without filtering, the code crashes after just a few
crossing times.  By filtering out the top $1/3$ spectral coefficients
as described above, the evolution can be extended as far as $t=60$.
No other choice of filtering seems to improve this further.  We also
run the test with an amplitude of $A=0.1$.  For this amplitude, the
evolutions are stable with filtering but unstable
without. Figs.~\ref{fig:ShiftedGaugeWaveCE}
and~\ref{fig:ShiftedGaugeWaveEE} show the constraints and error energy
for these evolutions.  The initial constraint violations in these
runs, $\|\mathcal{C}\| \approx 10^{-13}$, are at the level of the
finite machine precision, so increasing the resolution causes
increased, rather than decreased, constraint violations.  The growth
in $\|\delta \mathcal{U}\|$ seen in Fig.~\ref{fig:ShiftedGaugeWaveEE}
is linear in time for $t< 100$, becoming quadratic at late times.  The
quadratic in time growth is dominated by time-stepping error, which
tests show is convergent.  (Reducing this error to the level of
spatial truncation error would require a prohibitive amount of
computing time at the higher resolutions.)


\section{Gowdy Spacetime}
\label{sec:GowdySpacetime}

The Gowdy spacetimes are dynamic cosmological solutions that present a
serious challenge to any numerical relativity code.  The Gowdy
spacetimes are vacuum cosmological models having two spatial Killing
fields (planar symmetry) that expand from (or, when time-reversed,
contract toward) a curvature singularity.  Two particular examples of
these spacetimes with relatively simple analytical forms were chosen
for the Mexico City tests: one in which the spacetime expands away
from the singularity; another in which it collapses toward the
singularity.

\subsection{Expanding Form}
\label{subsec:GowdyExpanding}

The metric chosen for the expanding case is
\begin{equation}
  ds^2 = t^{-1/2}e^{\frac{\lambda-\lambda_0}{2}}(-dt^2 + dz^2) 
  + t(e^{P}dx^2 + e^{-P}dy^2)\ ,
  \label{eq:GowdyMetric}
\end{equation}
where
\begin{equation}
  P(t,z) = J_0(2\pi t) \cos(2\pi z)\ ,
\end{equation}
\begin{eqnarray}
  \lambda(t,z) &=& -2\pi t J_0(2\pi t) J_1(2\pi t)\cos^2(2\pi z)
  \nonumber \\ &&
  + 2\pi^2 t^2 \left[ J_0^2(2\pi t) + J_1^2(2\pi t) \right],
\end{eqnarray}
$\lambda_0 = \lambda(1,1/8)$, and $J_n$ is a Bessel function.
Asymptotically, $P$ approaches zero as time increases, and $\lambda$
increases linearly with time.  Because the metric components are
singular at $t=0$, the Mexico City test begins the evolution at $t=1$
and proceeds forward in time.

\begin{figure}
  \begin{center}
    \includegraphics[width=\linewidth, clip=false, trim= 0 0 0 0]%
    {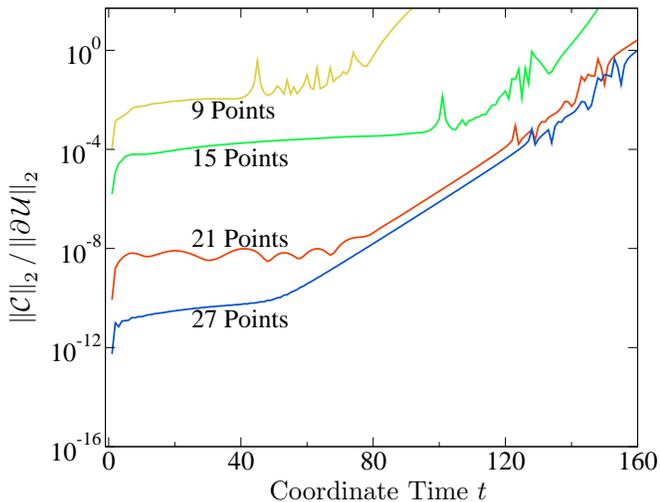}
  \caption{\textbf{Constraints for Expanding Gowdy Spacetime.}
    At early times, the exponential convergence of spectral methods 
    is clearly visible.  Soon, however, the evolutions are dominated 
    by constraints growing roughly as $e^{t/5}$.
    \label{fig:GowdyExpandingNormalizedKstCE}}
\end{center}
\end{figure}

\begin{figure}
  \begin{center}
    \includegraphics[width=\linewidth, clip=false, trim= 0 0 0 0]%
    {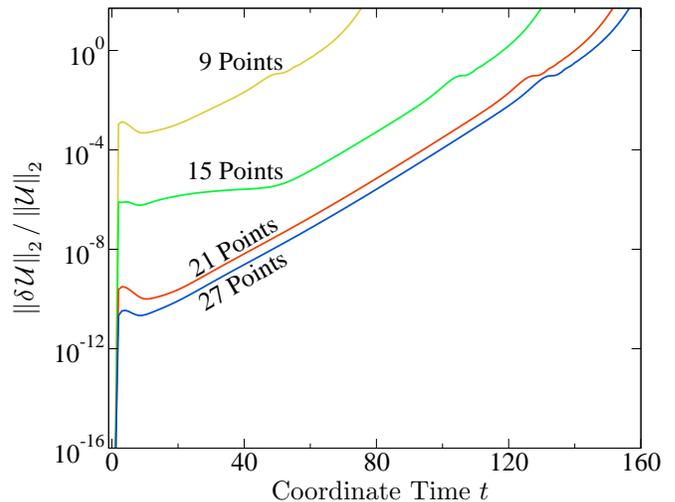}
  \caption{\textbf{Error Energy for Expanding Gowdy Spacetime.}
    The error energy converges with increased spatial resolution, but
    $\|\delta \mathcal{U}\|_2/\|\mathcal{U}\|_2$ grows like $e^{t/5}$.
    \label{fig:GowdyExpandingNormalizedKstE}}
\end{center}
\end{figure}

The time step $\Delta t$ required for numerical stability is roughly
given by the Courant condition $\Delta t\lesssim \Delta x/v$, where
$\Delta x$ is the spacing between collocation points and $v$ is the
coordinate speed of wave propagation, which in this case is the
coordinate speed of light. For the Gowdy metric the coordinate speed
of light in the $z$-direction is constant in time, but in the $x$- and
$y$-directions it varies roughly like $t^{3/4} e^{t/2}$.  Therefore,
the maximum allowed time step decreases in time like $t^{-3/4}
e^{-t/2}$, so for any fixed time step, the evolution will soon become
numerically unstable if there is any perturbation in the $x$- or
$y$-directions.  This problem can be circumvented by running the
simulation with just one point in the transverse directions,
effectively eliminating any perturbation that could seed the
instability.

Another difficulty with evolving the expanding Gowdy metric is that
the metric components and derivatives become enormous very quickly.
By $t \sim 725$ the numbers become larger than $10^{310}$, so the
evolution cannot be easily handled using standard 64-bit
floating-point arithmetic.  Our evolutions do not actually crash until
$t=700$; unfortunately errors dominate our evolutions long before this
time, as seen in Fig.~\ref{fig:GowdyExpandingNormalizedKstE}.  The
normalized error energy---along with the constraints shown in
Fig. 12---grows roughly as $e^{t/5}$, and accuracy is completely lost
in these evolutions by $t\sim 150$.

\subsection{Collapsing Form}
\label{subsec:GowdyCollapsing}

The time coordinate in the Gowdy metric given above can be transformed
so that the initial singularity is approached only asymptotically in
the past.  The new time coordinate, $\tau$, is defined by \mbox{$\tau
  \equiv \frac{1}{c}\, \ln \left( {t}/{k} \right)$}, where \mbox{$c =
  0.0021195119214607454$}, and \mbox{$k = 9.6707698127640558$}.  The
spacetime can be evolved backwards indefinitely without reaching the
singularity; that is, the time step is chosen to be negative.  For
purposes of convenience, the evolution is begun at an initial time of
$\tau = \tau(t_0)$, where $t_0 = 9.8753205829098263$, which is a zero
of $J_0(2\pi t)$.

\begin{figure}
  \begin{center}
    \includegraphics[width=\linewidth, clip=false, trim= 0 0 0 0]%
    {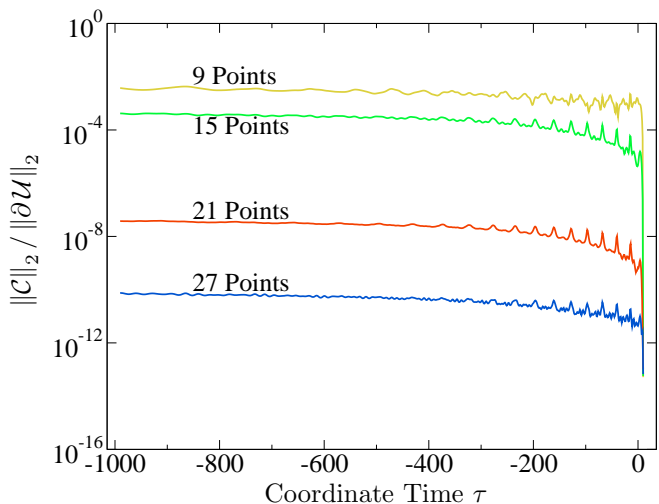}
  \caption{\textbf{Constraints for Collapsing Gowdy Spacetime.}  
    Note that the simulation starts at $\tau \sim 9.875$, and proceeds
    backwards.
    \label{fig:GowdyCollapsingNormalizedKstCE}}
\end{center}
\end{figure}

\begin{figure}
  \begin{center}
    \includegraphics[width=\linewidth, clip=false, trim= 0 0 0 0]%
    {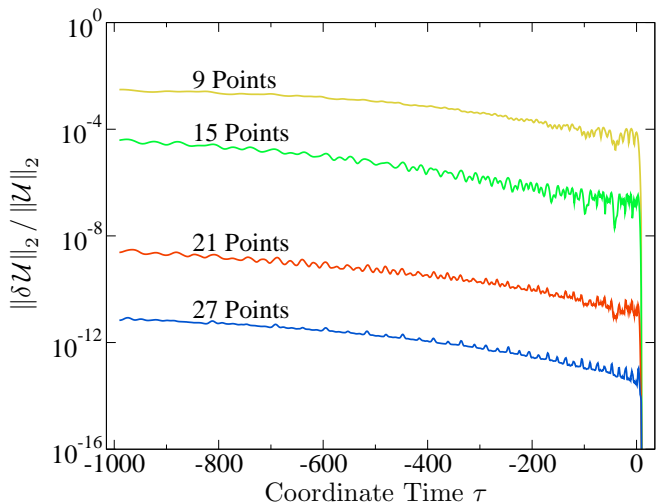}
  \caption{\textbf{Error Energy for Collapsing Gowdy Spacetime.} 
    The simulation starts at $\tau \sim 9.875$, and proceeds 
    backwards.
    \label{fig:GowdyCollapsingNormalizedKstE}}
\end{center}
\end{figure}

This evolution is far less challenging than the expanding case.  This
is because the lapse function is essentially an exponential in $\tau$,
so that the spacetime is becoming \emph{less} dynamical as the
simulation progresses and $\tau$ becomes more negative.  The main
challenge in this test is resolving the spatial features of the
solution.  For spectral methods, the convergence should be exponential
with increasing resolution, which is indeed the behavior shown in
Figs.~\ref{fig:GowdyCollapsingNormalizedKstCE}
and~\ref{fig:GowdyCollapsingNormalizedKstE}.


\section{Discussion}
\label{sec:conclusion}
We have applied the full suite of Mexico City
tests~\cite{Alcubierre2004}---modified suitably---to a pseudospectral
implementation of the KST formulation of Einstein's equations.  We
have also implemented the shifted gauge-wave test suggested by Babiuc,
\textit{et al.}~\cite{Babiuc2006}, and suggested a number of minor
changes to the tests that make them better challenges for
pseudospectral methods.  These tests reveal that the KST equations
with pseudospectral methods demonstrate excellent convergence and
accuracy, along with very good stability in all but a few cases.  We
have derived a fundamental limit Eq.~(\ref{eq:Error}) for the time
step accuracy possible in a method-of-lines numerical simulation, and
have shown that our implementation is capable of quickly achieving
that limit in the simple case of a sinusoidal linear wave.  We have
also shown that the use of filtering is very effective in reducing
nonlinear aliasing instabilities.

The Mexico City tests provide a basic set of benchmarks for evaluating
any numerical relativity code: allowing direct comparisons between
different codes that use different numerical techniques and different
formulations of the Einstein Equations.  However, the tests in their
present form make too many implicit assumptions about the evolution
system and the numerical methods.  Since the creation of the tests,
numerical relativity codes have become more diverse: using a variety
of improved numerical techniques (fixed and adaptive mesh refinement,
higher order finite-differencing, multi-block methods, spectral
methods) and at least two evolution systems (generalized harmonic and
BSSN) capable of successfully evolving binary black hole spacetimes.

To accommodate the wide range of numerical methods and evolution
systems now being used, future tests need to be formulated in more
abstract terms.  We recommend the following specific changes to the
statement of the tests:

\begin{figure}
  \begin{center}
    \includegraphics[width=\linewidth, clip=false, trim= 0 0 0 0]
    {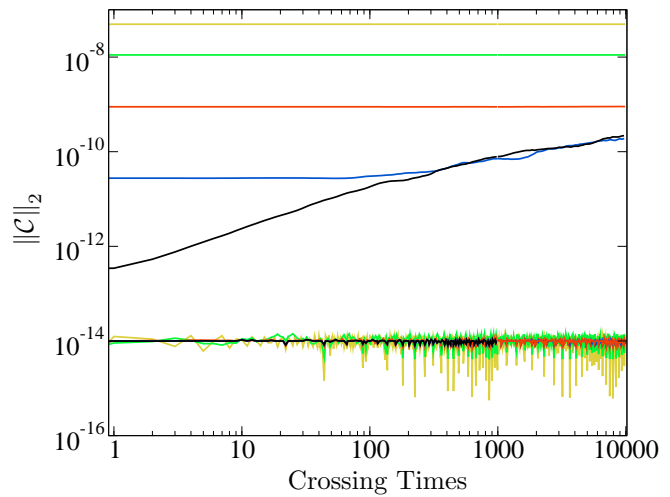}
    \caption{\textbf{Comparing Constraints for 1-D Gaussian Linear
        Waves}.  Hamiltonian constraint norms (ragged curves) are much
      smaller than $\|\mathcal{C}\|$ for this test, so by themselves
      are not good diagnostics of constraint violations.
      \label{fig:HamMomConstraint}}
  \end{center}
\end{figure}

\begin{enumerate}
\item A code should demonstrate convergence, both spatial and
  temporal, appropriate for the numerical method used, for each of the
  tests (gauge wave, linear wave, Gowdy spacetime, etc.).
\end{enumerate}
\noindent
The number of grid points or the time step needed to achieve a given
accuracy is highly dependent on the numerical implementation.
Therefore, the test specifications should not dictate a certain number
of grid points or a certain time-step size as the original formulation
of the tests did.
\begin{enumerate}
  \setcounter{enumi}{1}
\item The combined error of all evolution variables, and the combined
  constraint violation (including all constraints of an evolution
  system), cf. Eqs.~(\ref{eq:C}) and~(\ref{eq:deltaU}), should be
  reported for each of the tests.
\end{enumerate}
\noindent
Prescriptions for examining errors of particular variables or
constraints, such as those given in the original Mexico City tests,
are not applicable to evolution systems that do not evolve those
particular variables or constraints (e.g., tetrad or generalized
harmonic evolution systems).  In addition such prescriptions may not
encompass all variables or constraints (as in the KST system), and may
therefore fail to detect errors that accumulate only in a subset of
the evolved variables.  To illustrate this point,
Fig.~\ref{fig:HamMomConstraint} shows both the total constraint energy
$\|\mathcal{C}\|$, and the Hamiltonian constraint for the Gaussian
linear wave (cf. Fig.~\ref{fig:LinearWaveGaussianCE}).  The
Hamiltonian constraint turns out to be anomalously small for the KST
system in this case, and so is not a good overall error indicator.
\begin{enumerate}
  \setcounter{enumi}{2}
\item Use periodic Gaussian wave spatial profiles in the linear and
  gauge wave tests.
\end{enumerate}
The sinusoidal spatial profiles specified in the original Mexico City
tests with periodic boundary conditions provide an artificial
advantage for spectral techniques.  Periodic Gaussian profiles are no
more difficult for finite difference codes, and provide a
significantly greater challenge for spectral methods.  Finally,
\begin{enumerate}
  \setcounter{enumi}{3}
\item Output data at generic times, not at integer multiples of the
  light-crossing time.
\end{enumerate}
Outputting data at exact integer multiples of the light-crossing time
significantly underestimates the errors in codes with very small
dissipation (such as spectral codes).

We believe these recommendations will make it easier to apply the
Mexico City tests fairly to a far wider class of numerical relativity
codes, and so facilitate apples-with-apples comparisons between these
codes.  We have learned a great deal about the subtle properties of
our code by carefully running and analyzing these simple tests.  We
encourage other groups to make their results from these tests public
so that meaningful and objective comparisons between codes can be
made.


\acknowledgments We thank Rob Owen, Luisa Buchman and Olivier Sarbach
for helpful conversations.  This work was supported in part by a grant
from the Sherman Fairchild Foundation to Caltech and Cornell; by NSF
grants PHY-0099568, PHY-0244906, DMS-0553302, PHY-0601459, and NASA
grants NAG5-12834 and NNG05GG52G at Caltech; and by NSF grants
PHY-0312072, PHY-0354631, and NASA grant NNG05GG51G at Cornell.


\bibliography{References}


\end{document}